\documentclass[pdflatex,sn-mathphys]{sn-jnl}

\jyear{2021}  %

\theoremstyle{thmstyleone}

\theoremstyle{thmstyletwo}

\theoremstyle{thmstylethree}

\begin{document}

\title[Stock and market index prediction using Informer network]{Stock and market index prediction using Informer network}

\author[1]{\fnm{Yuze} \sur{Lu}}\email{123zzzzlong123@gmail.com}
\equalcont{These authors contributed equally to this work.}

\author[1]{\fnm{Hailong} \sur{Zhang}}\email{yuze.luu@gmail.com}
\equalcont{These authors contributed equally to this work.}

\author*[1]{\fnm{Qiwen} \sur{Guo}}\email{gqw995@gmail.com}

\affil[1]{\orgdiv{Shenzhen Ping An Integrated Financial Services Co., Ltd. Shanghai Subsidiary}, \orgaddress{\city{Shanghai} \postcode{201210}, \country{China}}}

\abstract{Applications of deep learning in financial market prediction has attracted huge attention from investors and researchers. In particular, intra-day prediction at the minute scale, the dramatically fluctuating volume and stock prices within short time periods have posed a great challenge for the convergence of networks result. Informer is a more novel network, improved on Transformer with smaller computational complexity, longer prediction length and global time stamp features. We have designed three experiments to compare Informer with the commonly used networks LSTM, Transformer and BERT on 1-minute and 5-minute frequencies for four different stocks/ market indices. The prediction results are measured by three evaluation criteria: MAE, RMSE and MAPE. Informer has obtained best performance among all the networks on every dataset. Network without the global time stamp mechanism has significantly lower prediction effect compared to the complete Informer; it is evident that this mechanism grants the time series to the characteristics and substantially improves the prediction accuracy of the networks. Finally, transfer learning capability experiment is conducted, Informer also achieves a good performance. Informer has good robustness and improved performance in market prediction, which can be exactly adapted to real trading.}


\maketitle

\section{Introduction}\label{sec1}
Stock market is one of the most active financial market activities around the world. Because of its large number of participants, driven by the flow of enormous wealth, abundant news and policy information, changes in the stock market have attracted the attention of both investors and researchers for decades. In the history of capital market, people have learned predict stock trend based on fundamental analysis \cite{bib1} and technical analysis \cite{bib2}. In recent years, with the rapid development of computer technology, it is easy for researchers to predict stock dynamic changes based on the huge amount of historical data \cite{bib3} and summarizing the characteristics of stock changes. More and more institutions and investors are trying to invest based on objective analysis given by computers, avoiding the influence of subjective thinking. Due to the high complexity of the stock market, feature extraction using deep learning has become the focus of research \cite{bib4}\cite{bib5}. With the support of big data, financial market prediction based on deep learning has become a hot approach to solve the problem.

For many value investors, fundamental analysis of companies is still the most common approach. The fundamental analysis considers. The company's financial position, assets and liabilities, industry outlook, company status and policy changes, and even certain unforeseen events, news reports and natural disasters \cite{bib6}\cite{bib7}\cite{bib8} are also taken into account. Despite the fact that these variables are crucial for predicting stock price movements and can affect a company's stock price in the medium and long term, they are unable to predict stock price fluctuations over short periods of time \cite{bib9}. But for most general investors and institutions, technical analysis is very popular. After the process of technical analysis, a number of intuitive indicators \cite{bib10}\cite{bib11} are generated to help people judge: moving average (MA), exponential moving average (EMA), moving average convergence/ divergence rules (MACD) and so on. However, the judgment of technical indicators is still mixed with a strong sense of subjectivity, making it impossible to be completely objective and thorough when trading. Moreover, according to \cite{bib12}\cite{bib13} technical analysis has a variety of limitations.

With the advancement of computer technology, particularly the quick rise of artificial intelligence, research on stock price forecasting is also becoming more abundant and in-depth. The most popular ones in recent years are RNN-based networks. Especially, Long Short-Term Memory network (LSTM) and its derivative networks are the most popular type of network in current stock price forecasting, market index forecasting, portfolio optimization and other areas of financial activities \cite{bib14}. The typical LSTM network consists of cell, input gate, an output gate, and forget gate \cite{bib5}. The unique structure provides it with the ability to store short-term memory and based on which, it can predict near future stock price trends and short-term stock price fluctuations. \cite{bib15} determined the accuracy of stock predict by LSTM, and reach to that changing training epoch can enhance the models performance. \cite{bib16} utilized the LSTM as encoder and decoder, and added the attention mechanism into model. \cite{bib17}\cite{bib18} built new LSTM-based networks to predict the opening and closing prices of several stocks by some indicators, respectively, and achieved good results. Among some combinatorial models or improved models, it also has excellent experimental results. \cite{bib19} created a new composite CNN-LSTM, they fed the input data sequentially into the convolutional and pooling layers in the CNN layer, extract features from the input data, and fed the extracted features further into the LSTM. They improved the accuracy by up to 2.2$\%$ on top of the comparison networks based on the Shanghai composite index. \cite{bib20} tested a new Attention-LSTM network, which was very novel to embed attention mechanism into the traditional LSTM. They also compared the Attention-LSTM with ARIMA, LSTM and Stacked-LSTM \cite{bib21}, the former was proved to be optimal and predicted financial time series much better. \cite{bib22} compared LSTM-based deep recurrent neural network (DRNN) with LSTM, and associated net on Shanghai composite index, PetroChina and ZTE, surprisingly finding that the LSTM performs poorly and the DRNN outperformed only in multi-target prediction task. \cite{bib23} executed a comparative experiment between the LSTM, SVM, CNN, NFNN and multiple pipeline models on S\&P 500 stock, used multiple LSTMs to learn the time dependence of different time scale features and combined all the information to predict the future closing price. They proved LSTM performed best in all networks. \cite{bib24} and \cite{bib25} also proved LSTM-based or improved LSTM models are the best choices for prediction tasks.

Even though LSTM has been demonstrated to be among the top network models in terms of overall performance in a number of related experiments, its serial network structure necessarily reveals several flaws in addition to providing good flexibility. The increased memory storage requirements and complicated back propagation procedure \cite{bib26}, which make the results scarcely converge, lead the inference speed to drop significantly as the input sequence lengthens and the MSE score to continually rise. 

Recent years, Transformer \cite{bib27} was developed to address the issue of serial computing, which causes a sharp decrease in speed. Transformer introduces a multi-headed self-attention mechanism and takes a parallel computing approach to reduce computation time and also performance degradation due to long-term dependencies. And in recent years, Transformer has gradually surpassed LSTM as the dominant method for handling and forecasting time series data \cite{bib28}. It is also becoming more prevalent in the area of stock prediction. \cite{bib29} predicted CSI 300, S\&P 500, Hang Seng Index, and Nikkei 225, using Transformer, B$\&$H, RNN, CNN and LSTM, and proved the superiority of Transformer. \cite{bib30} built a Transformer-based network, made full use of the self-attention mechanism on 12 stocks, and significantly increase investment accuracy. \cite{bib31} utilized Transformer’s multi-head self-attention, ran on the hourly data on the Dogecoin and obtained a prediction accuracy of 98.46$\%$ and R-squared value of 0.8616. But according to the experimental feedback, Transformer didn’t perform well where stock price fluctuated dramatically. \cite{bib32} created a new network DTML based on Transformer, and predicted six market indices from different countries. They got up to 13.8$\%$ higher profits than compared LSTM-like networks. Meanwhile, some studies have also tried to use Transformer to assist in predicting stock prices based on news media information. \cite{bib33} used LTN with local constraints to encode financial news, which also could increase the attention weight of key local information. \cite{bib34} also studied that use BERT to make forecasts based on investor sentiment and technical analysis, while they also wanted more researches to focus on intraday trading.

Experiments demonstrated that the Transformer-based networks are seemly superior to LSTM-based ones in stock price prediction. Nevertheless, Transformer-based models research for financial market prediction is still not deep enough; there is a significant research lag in financial area and natural language processing (NLP). Despite the excellent works mentioned, with the increase of the input sequence, the disadvantages and challenges of Transformer are still unsolved:

\begin{itemize}

\item The time complexity and memory usage per layer to be $\mathcal{O}(L^2)$, where L is the length of sequence.

\item The memory bottleneck in stacking layers for long in-puts, which limits the model scalability in receiving long sequence inputs.

\item The step-by-step inference in vanilla Transformer decoding makes the speed plunge in predicting long outputs.

\end{itemize}

The improved model like Sparse Transformer \cite{bib35}, LogSparse Transformer \cite{bib36} and Longformer \cite{bib37} could only improve performance to a very limited extent. The forecasting model's capacity to predict changes in stock changes, especially at the minute level in real time, is severely constrained, particularly by the third limitation. As mentioned before, with the input sequence getting longer, the MSE score keeps rising, which causes the model to run into the issue that all LSTM-like models encounter: the longer the sequence, the more likely the results will diverge. In addition, the existing Transformer-based networks decoders output results serially, leading to inefficiency. However, as the focus of this research is minute-level dynamic stock forecasting, extended forecasting times and subpar outcomes must have a variety of detrimental implications on following trading.

Necessarily, in order to solve the above problems, in this paper we introduce Informer \cite{bib38} into the intraday stock price prediction task. Informer has won the Outstanding and Distinguished Papers awards in AAAI 2021. It was proposed based on Transformer network. It well solves the above three key problems faced by Transformer, reduces computing time and accomplishes the prediction function of long time sequences. These advantages of Informer are perfectly suited to the task of intraday long sequence stock price rapid prediction.

\section{Methodologies}\label{sec2}

\subsection{Proposed network}\label{subsec1}

Informer has added three significant changes and a few minor ones to Transformer in order to maintain strengths and change weaknesses of Transformer. The main structure of Informer and Transformer are shown in Fig. 1:

\begin{figure}[ht]%
\centering
\includegraphics[width=0.9\textwidth]{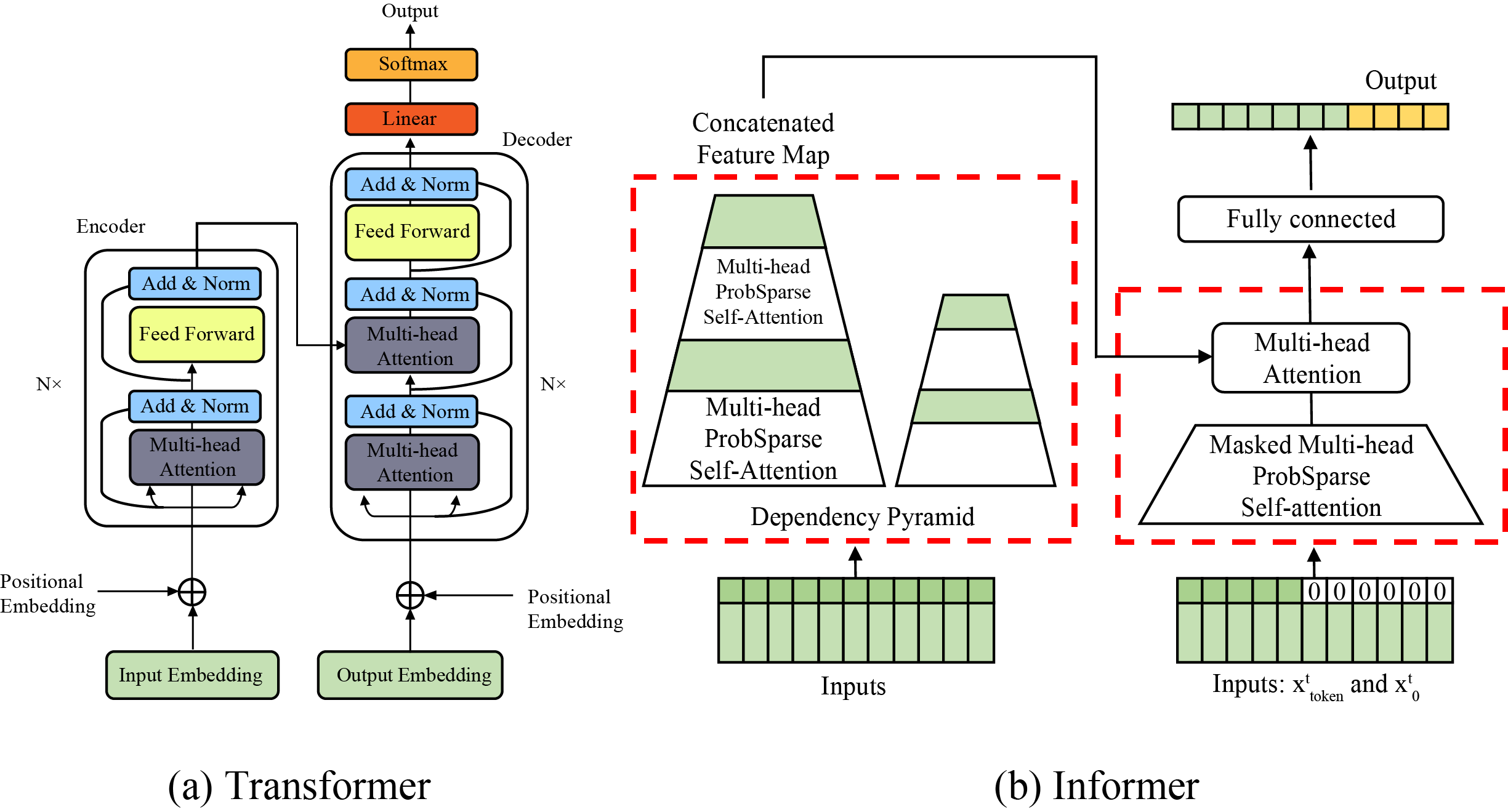}
\caption{The overall architecture of Transformer and Informer}\label{fig1}
\end{figure}

\subsubsection{ProbSparse self-attention mechanism}\label{subsubsec1}

In the original Transformer, the self-attention mechanism calculating inputs, Query $(Q \in \mathbb R^{L_Q \times d})$ , Key $(K \in \mathbb R^{L_K \times d})$ and Value $(V \in \mathbb R^{L_V \times d})$ \cite{bib27}, consumes a large amount of computing resources. The probability distribution of self-attention, however, exhibits potential sparsity in probability studies, indicating that the majority of the attention has a relatively weak role and that only a small number of dot-product pairings significantly influence the main attention. ProbSparse self-attention mechanism is proposed that the i-th Query’s attention is deﬁned in a probability form: 

\begin{equation}
A(q_i,K,V)=\sum\limits_{j} \frac {k(q_i,k_i)}{\sum\limits_{l}k(q_l,k_l)}v_i=E_{p(k_i \lvert q_i)}[V_j],\label{eq1}
\end{equation}

\noindent where $p(k_j \lvert q_j)=k(q_i,k_j)/\sum\limits_{l}k(q_i,k_l)$ and $k(q_i,k_j)$ selects the asymmetric exponential kernel $exp(q_ik_j/\sqrt {d})$. To measure the sparsity of the query, Kullback-Leibler divergence is used to calculate the relative entropy of the attention probability distribution of the Query versus the uniformly distributed probability distribution, where the evaluation formula for the sparsity of the i-th Query is:

\begin{equation}
M(q_i,K)=ln\sum\limits_{j=1}^{L_K} e^{\frac {q_ik_j}{\sqrt{d}}}-\frac{1}{L_K}\sum\limits_{j=1}^{L_K} \frac{q_ik_j}{\sqrt{d}},\label{eq2} 
\end{equation}

\noindent where $ln\sum\limits_{j=1}^{L_K} e^{\frac {q_ik_j}{\sqrt{d}}}$ is the Lg-Sum-Exp (LSE) of $q_i$  on all the Keys, and $\frac{1}{L_K}\sum\limits_{j=1}^{L_K} \frac{q_ik_j}{\sqrt{d}}$ is the arithmetic mean. The final ProbSparse self-attention formula is:

\begin{equation}
A(Q,K,V)=\emph{Softmax}(\frac {\overline{Q}K}{\sqrt{d}})V,\label{eq3}
\end{equation}

\noindent where $\overline{Q}$ is the sparse matrix of the same size of $q$ and it only contains the top-u Queries under the sparsity measurement $M(q,K)$. Through the above steps, “Lazy” Query can be eliminated and “Active” ones are maintained; thus, space complexity is reduced from $\mathcal{O}(L^2)$ to $\mathcal{O}(LlnL)$. Computational complexity of the quadratic in original network is solved effectively.

\subsubsection{Self-attention distilling operation}\label{subsubsec2}

A self-attention distilling operation is carried out in encoder part. Encoder is a structure set up to extract the dependencies of long sequential inputs in the far future, which is also a result of ProbSparse self-attention mechanism. The feature mapping of encoder has redundant combinations of Values. Therefore, using the distilling operation is to assign higher weights to dominant features with dominant features and generate at the next level focus self-attention feature mapping. Distilling operation from j-th layer to j+1-th layer is:

\begin{equation}
X^t_{j+1}=\emph{Maxpool}(\emph{ELU}(\emph{Conv1d}([X_j^t]_{AB}))),\label{eq4}
\end{equation}

\noindent where $[X_j^t]_{AB}$ contains the key operations in attention blocks and multi-head \cite{bib28} ProbSparse self-attention; \emph{Conv1d} represents a one-dimensional convolution operation on a time series; \emph{ELU} is activation function. The distillation mechanism allows each layer of the decoder to reduce the length of the input sequence by half, thus greatly saving the memory cost and computation time. The detailed distillation process is shown in Fig. 2.

\begin{figure}[ht]%
\centering
\includegraphics[width=0.9\textwidth]{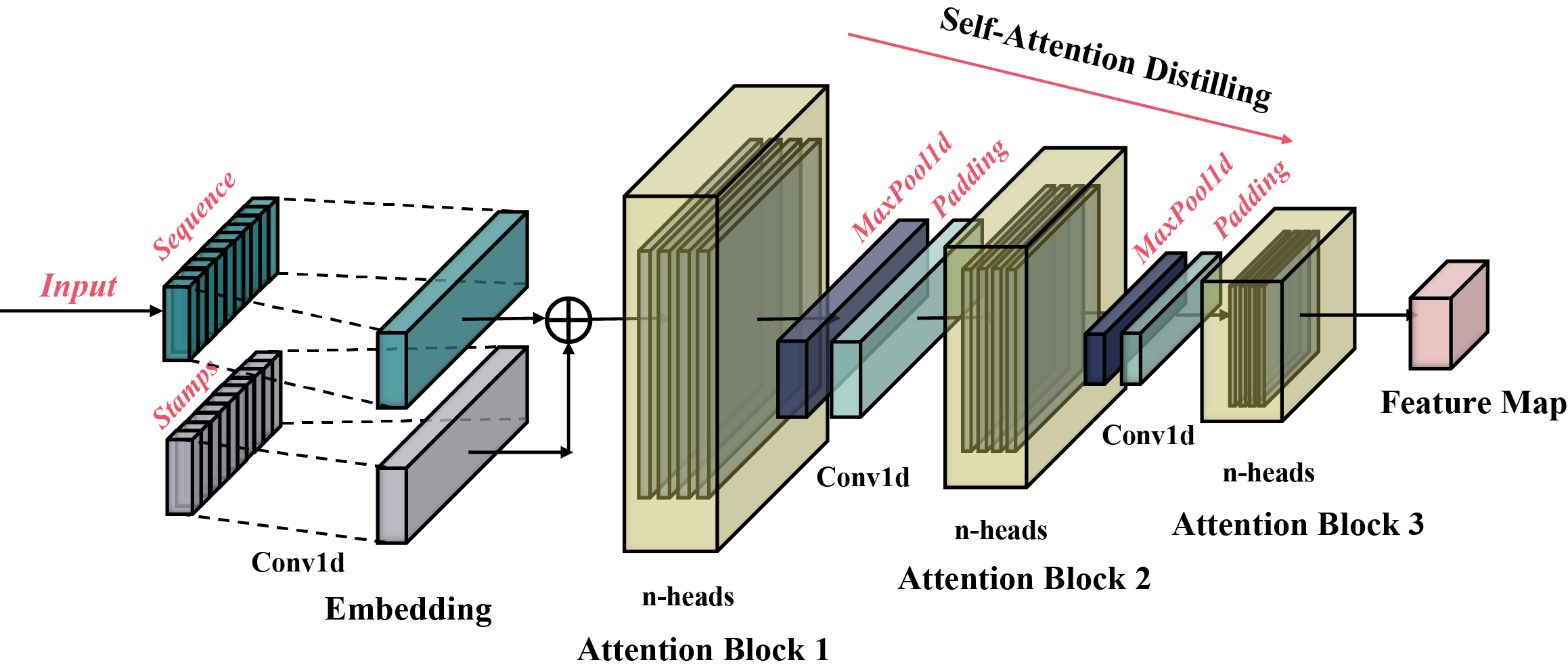}
\caption{Distillation process of Informer}\label{fig2}
\end{figure}

\subsubsection{Generative inference method}\label{subsubsec3}

To cope with the prediction of long sequence problem, based on a standard decoder structure \cite{bib27}, the generative style inference method is employed to reduce the long-term prediction's speed drop. The decoder is fed with the following vectors:

\begin{equation}
X^t_{de}=\emph{Concat}(X^t_{token},X^t_0) \in (L_{token}+L_y) \times d_{model},\label{eq5}
\end{equation}

\noindent where $X^t_{token} \in L_{token} \times d_{model}$ is known guiding sequence, $X^t_0 \in L_y \times d_{model}$ s the sequence to be predicted. This decoder style can output all $X^t_0$ at once based on the known guiding sequence; instead of cascading dynamically and inefficiently like the RNN-based model. This new type of inference and output is real-time, more intuitive, making it easier to subsequently construct trading strategies. The output parts of traditional Transformer and Informer are shown in Fig. 3.

\begin{figure}[ht]%
\centering
\includegraphics[width=0.9\textwidth]{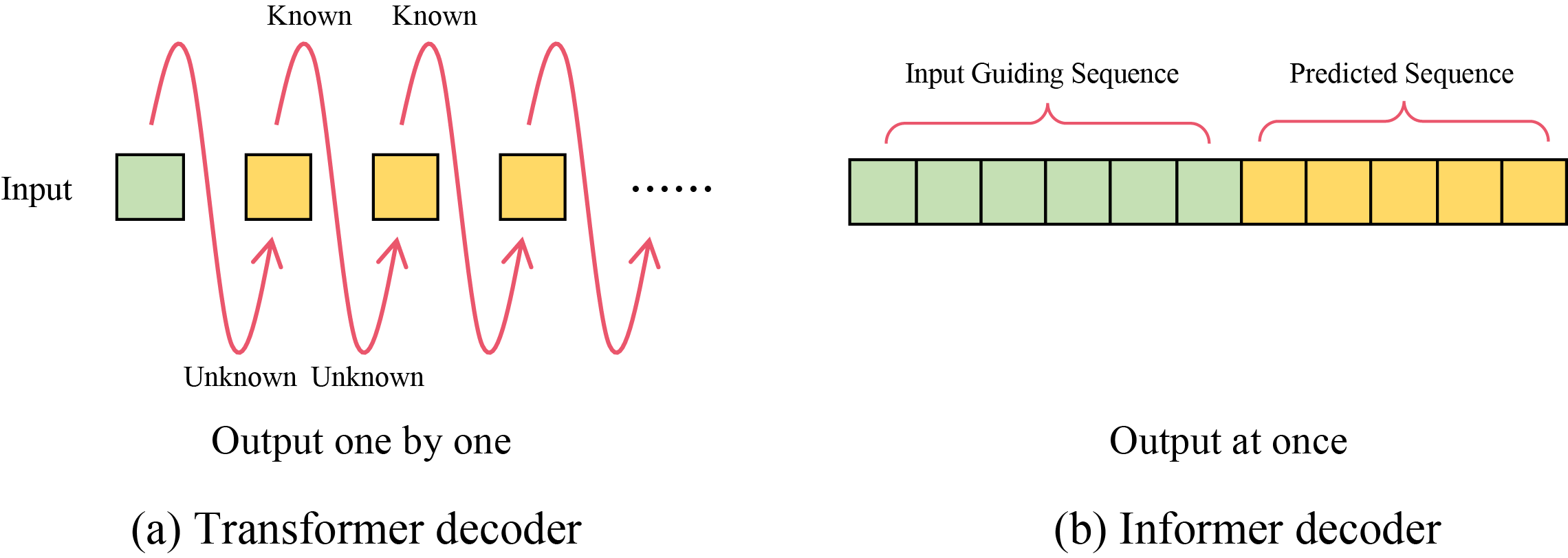}
\caption{Comparison of output methods traditional Transformer decoder (left) versus Informer decoder (right)}\label{fig3}
\end{figure}

\subsubsection{Global time stamp}\label{subsubsec4}

Transformer employs multi-head self-attention mechanism, thus avoiding the recursive method of RNN and speeding up the training time. However, the model itself does not have any sense of location when each sequence simultaneously passes through the encoder and decoder stack of the Transformer. Therefore, positional embedding \cite{bib39} method is needed to incorporate the sequential information of inputs into the model. 

Based on positional embedding, Informer introduces the global time stamp embedding operation, making the timestamp one of the characteristics as well. The time stamp section includes year, month, week, hour, and minute labels. By embedding a time stamp, the model is able to accurately match a minute of data to a certain hour of the day, giving each sequence a unique time characteristic. The stamped sequences are the fed into decoder. The detailed processing is shown in Fig. 4.

\begin{figure}[ht]%
\centering
\includegraphics[width=0.9\textwidth]{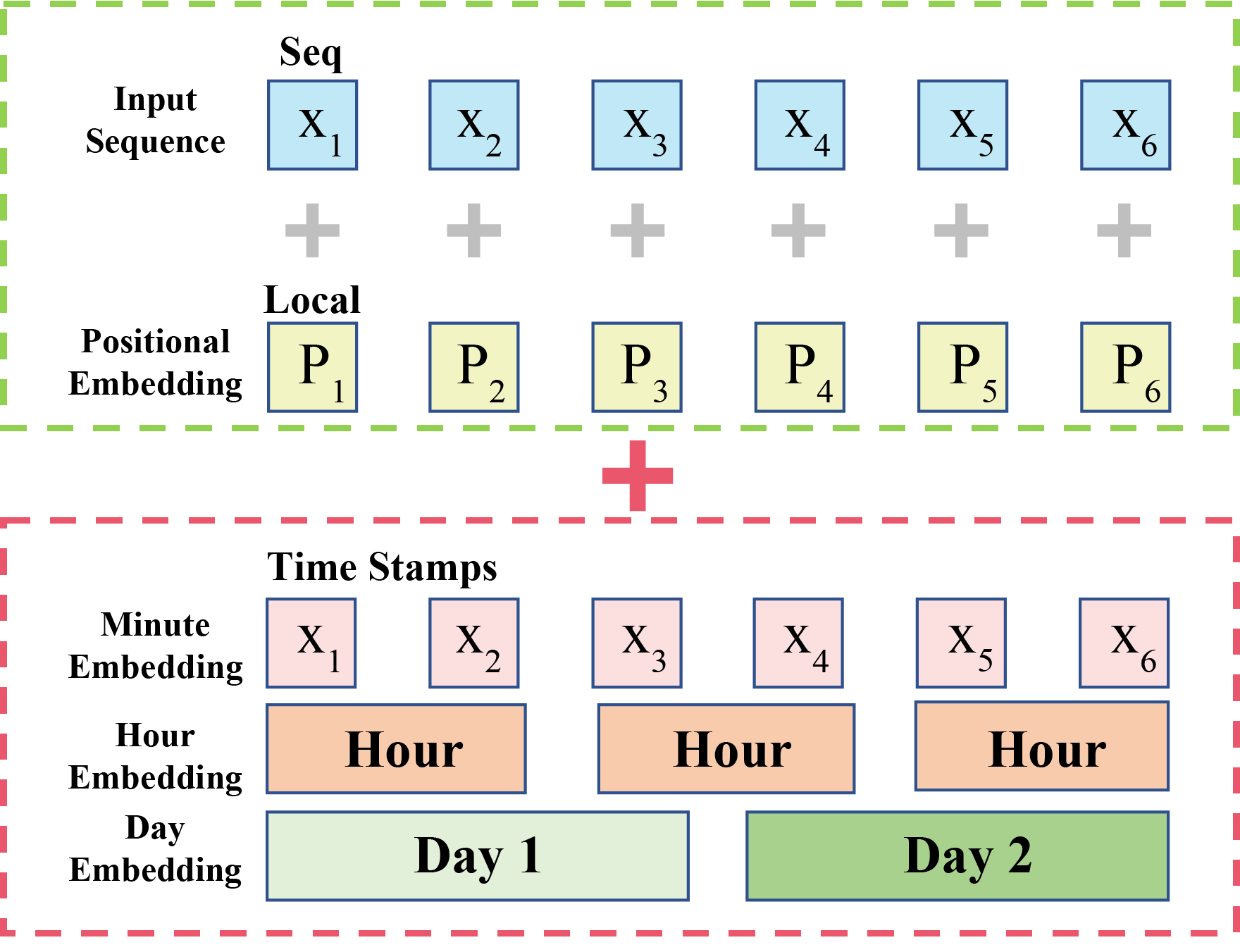}
\caption{Global time stamp processing in decoder of Informer}\label{fig4}
\end{figure}

Contrary to typical studies focused on day level changes, whose volume varies smoothly \cite{bib40}, this research focuses on dynamic forecasting on stock fluctuations at the minute level, where intraday volume fluctuations are quite violent. Due to the quick opening and closing of positions from numerous institutions and investors, the volume frequently increases multiple times \cite{bib41} during the first and last minutes of the market. The model may be at risk of gradient explosion \cite{bib42} because to the sudden volume increase, and may also be treated as extreme by the model, significantly affecting stock price forecasts and causing an increase in local errors. However, the volume at the start and end of each trading day determines the tone for the entire day as well as the trend for the following trading day. Additionally, erratic volume frequently causes sharp changes in stock prices. Making an accurate prediction can reduce numerous financial risks and lead to big profit opportunities. By putting special time stamps on the open and close data, the model can treat these data differently, thus avoiding unnecessary errors.

\subsection{Stock data acquisition}\label{subsec2}

In this work, two representative stock price indices and two typical company stocks are selected: Hong Kong Hang Seng Index (HSI), NASDAQ Index (IXIC), Tencent Holdings Ltd and Apple Inc. (AAPL). The data involved in the training includes opening, closing, high and low prices, volume and turnover ratio at 1-minute and 5-minute frequencies. The closing price is defined as the predicted target, and the rest of data labels are training features. The complete forecasting task is to predict univariate data from multivariate data.

The 1-minute data time period is from 03 January, 2022 to 12 December, 2022, and the 5-minute data time period is from 04 April, 2021 to 12 December, 2022. Among the whole data set, 70$\%$ of the data is set as training set to train the initial network; 10$\%$ of the data is set as validation set to monitor if the training process is going in a good direction; 20$\%$ of the data is defined as testing set to examine the performance of the proposed method under the performance evaluation criteria. Table 1 provides the detailed divisions of the stock data information:

\begin{table}[!ht]
  \centering
\begin{center}           
\begin{minipage}{\textwidth}
\caption{The datasets of the selected stock price indices and company stocks}\label{tab1}%

\resizebox{\linewidth}{!}{
\begin{tabular}{ccccccc}
\toprule
\multicolumn{2}{c}{Index and Stock} & Time period & Total time point & Training & Validation & Testing \\
\midrule
\multirow{2}{*}{HSI} & 1-minute & 2022/01/03-2022/12/12 & 65358 & 45751 & 6536 & 13072 \\
\multirow{2}{*}{ } & 5-minute & 2021/04/05-2022/12/12 & 28248 & 19774 & 2825 & 5650 \\
\cline{1-7}
\multirow{2}{*}{IXIC} & 1-minute & 2022/01/03-2022/12/12 & 77264 & 54085 & 7726 & 15453 \\
\multirow{2}{*}{ } & 5-minute & 2021/04/05-2022/12/12 & 33384 & 23369 & 3338 & 6677 \\
\cline{1-7}
\multirow{2}{*}{Tencent} & 1-minute & 2022/01/03-2022/12/12 & 65359 & 37848 & 5407 & 10814 \\
\multirow{2}{*}{ } & 5-minute & 2021/04/05-2022/12/12 & 28248 & 16356 & 2337 & 4673 \\
\cline{1-7}
\multirow{2}{*}{AAPL} & 1-minute & 2022/01/03-2022/12/12 & 77212 & 54048 & 7721 & 15442 \\
\multirow{2}{*}{ } & 5-minute & 2021/04/05-2022/12/12 & 33365 & 23356 & 3337 & 6673 \\

\botrule
\end{tabular}
} 
\end{minipage}
\end{center}
\end{table}

\section{Experiment and performance evaluation}\label{sec3}

\subsection{Network hyper-parameters}\label{subsec1}

In this experiment, in addition to the Informer network to be evaluated, the compared networks are included the vanilla LSTM \cite{bib43}, the conventional Transformer \cite{bib27} and the BERT network \cite{bib44}. After implementing serval experiments in advance, the ideal hyper-parameters are identified to ensure the optimization of the training process. The detailed parameters are shown in Table 2.

\begin{table}[ht]
\begin{center}
\begin{minipage}{\textwidth}
\caption{Key hyper-parameters in Informer}\label{tab2}%
\resizebox{\linewidth}{!}{
\begin{tabular}{cc|cc}
\toprule
Parameter & Value  & Parameter & Value \\
\midrule
Batch size & 32 & Dropout & 0.05  \\
Loss function & MSE & Epochs & 20  \\
Input sequence & 96 & Early stopping patience & 3  \\
Guiding sequence & 48 & Leaning rate & $4 \times 10^{-5}$  \\
Predicted sequence & 24 & Activation & GeLU \\
\botrule
\end{tabular}
}
\footnotetext{Note: Dropout and early stopping patience mechanism is to prevent overfitting; early stopping is controlled by validation loss; learning rate is dynamic, gradually growing smaller with the training process.}
\end{minipage}
\end{center}
\end{table}

The other comparative network parameters are defined properly, referred and replicated from \cite{bib45}\cite{bib46}\cite{bib47}. During training process, the loss functions of all models have shown distinct downward trend, converged and stopped early after sufficient training.

The experiments are carried out on a Windows PC with an Inter Core i7-10750H CPU, RTX2070 GPU, and programming with Python 3.7 and Pytorch.

\subsection{Performance evaluation criteria}\label{subsec2}

The prediction performance of the models is compared and evaluated by the predicted and ground truth data of the output. In this experiment, the networks are evaluated from various perspectives using three evaluation criteria. The data involved in the measurement are from the testing sets of experimental models.

\begin{itemize}

\item Mean Absolute Error (MAE):

\begin{equation}
MAE=\frac{1}{N}\sum\limits_{i=1}^{N} \lvert \hat{x}_i-x_i \rvert,\label{eq6}
\end{equation}

where, $hat{x}_i$ is the predicted value, $x_i$ is the ground truth, and $N$ is the number of the whole sequence. The MAE criteria can prevent errors from canceling one other out by averaging the absolute errors between predicted and true values. MAE is a linear score in which all individual differences are equally weighted on the mean. It is utilized widely in many experiments \cite{bib48}\cite{bib49}.

\item Root Mean Square Error (RMSE):

\begin{equation}
RMSE=\sqrt{\frac{1}{N} \sum\limits_{i=1}^{N} {(\hat{x}_i-x_i)}^2},\label{eq7}
\end{equation}

where $\hat{x}_i$ is the predicted value, $x_i$ is the ground truth, and $N$ is the number of the whole sequence. RMSE is used to measure the deviation between the predicted and true values, and it is more sensitive to abnormal values. When there are extreme values, it can strongly affect the value of RMSE, and there are generally higher RMSE values in unstable, or active trading times \cite{bib50}\cite{bib51}.

\item Mean Absolute Percentage Error (MAPE):

\begin{equation}
MAPE=\frac{1}{N} \sum\limits_{i=1}^{N} \lvert \frac{\hat{x}_i-x_i}{x_i} \rvert \times 100
\end{equation}

\noindent where $\hat{x}_i$ is the predicted value, $x_i$ is the ground truth, and $N$ is the number of the whole sequence. MAPE is a relative error measure that uses absolute values to avoid positive and negative errors from canceling each other out and is often used to compare the accuracy of various time series models for forecasting \cite{bib52}\cite{bib53}.

\end{itemize}

In the experiments, the smaller the values of the above three evaluation criteria, the smaller the deviation of the predicted value from the true value, and the better the model performance. In case of conflicting results between criteria, MAPE is used as the main referred standard \cite{bib54}.

\subsection{Prediction experiment on the minute scale}\label{subsec3}

The experiments are conducted on datasets: Hong Kong Hang Seng Index (HSI), NASDAQ Index (IXIC), Tencent Holdings Ltd and Apple Inc. (AAPL). We have carried out experiments on different time scales, including 5-minute scale and 1-minute scale, which means that there are 8 data sets to be predicted. The networks participating in the experiment are LSTM, Transformer, BERT and Informer, and like mentioned previously, hyper-parameters are set properly to guarantee that all the networks reach to convergence. The evaluation criteria will separately measure the prediction performance of different networks in different tasks. In order to display the results more clearly, 250 continuous points are randomly selected for each group of experimental outputs; the ground truth, the predicted values of the different networks and the volume corresponding to time points are shown in Fig. 5 and Fig. 6.

\begin{figure}[ht]%
\centering
\includegraphics[width=0.9\textwidth]{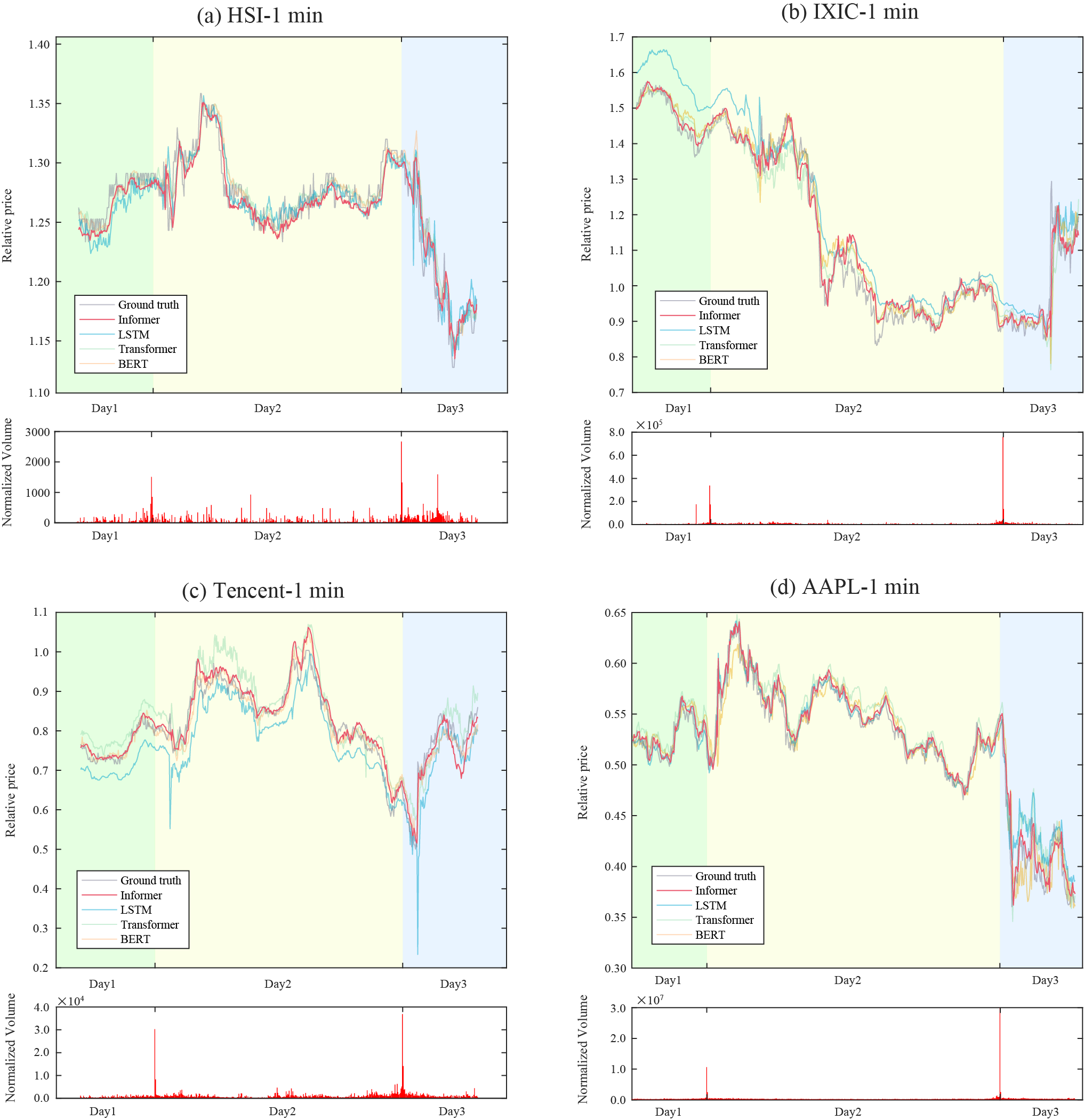}
\caption{Informer and compared networks prediction performance on 1-minute datasets}\label{fig5}
\end{figure}

\begin{figure}[ht]%
\centering
\includegraphics[width=0.9\textwidth]{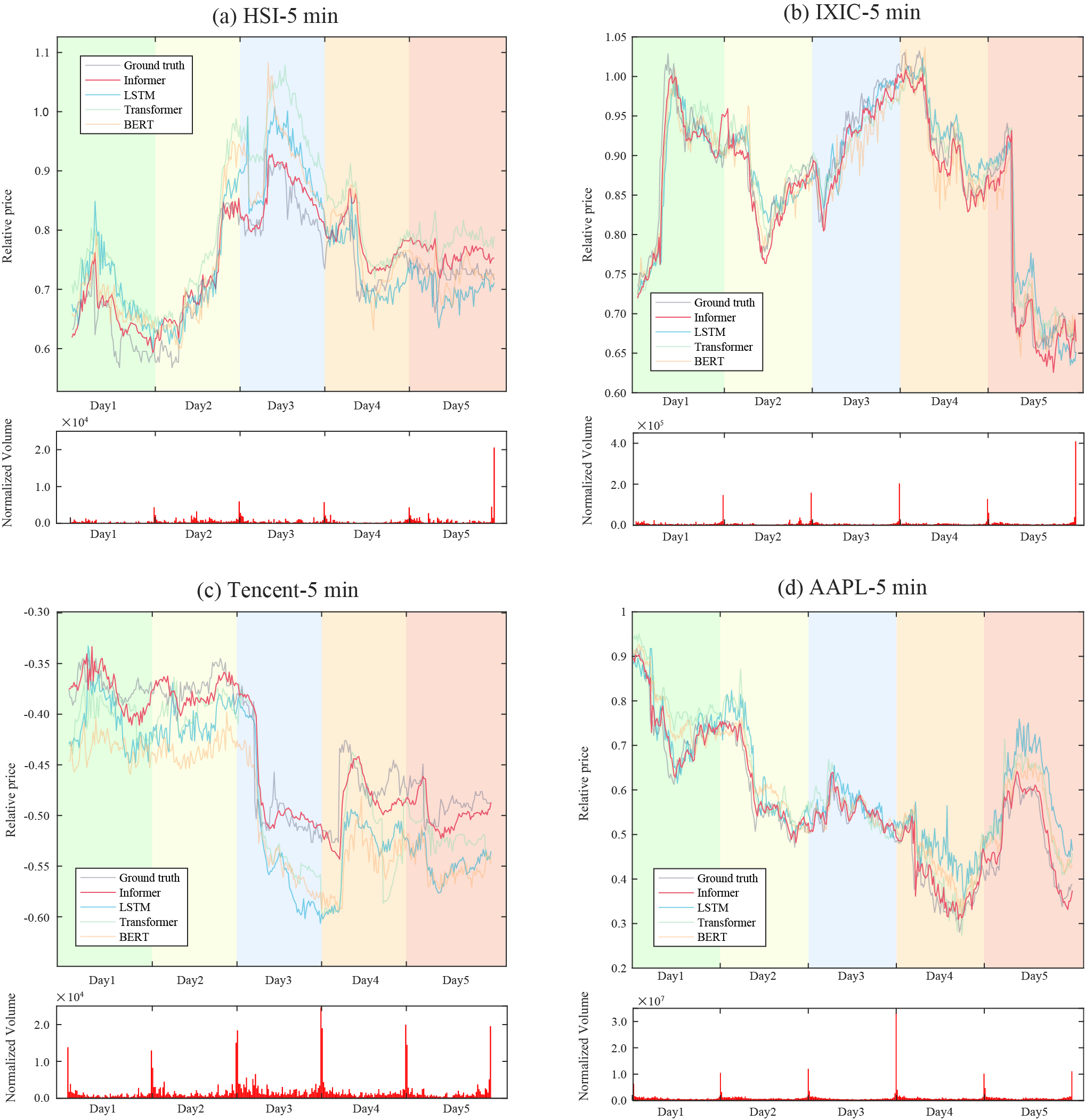}
\caption{Informer and compared networks prediction performance on 5-mintue datasets}\label{fig6}
\end{figure}

Intuitively, among all the experimental networks, the LSTM network output has the largest overall deviation of the prediction results; it can predict the movement trend of stock prices well, but there will be a marked lag, which is mainly affected by the structure of its serial output structure, causing the real-time performance to be poor. And LSTM is not sensitive for large fluctuations within a short time period. Next, Transformer outputs predictions with obviously less lag, but at relatively flat stock price positions, Transformer's predictions show more noise; besides, the process is more complicated than LSTM in training part. BERT is the second most effective among all networks and can restore the ground truth to a large extent; the error between the prediction result and the ground truth is also acceptable. However, for one of the key concerns of this paper, namely the performance at active trading periods, all three networks above show significant and substantial deviations and errors. Plainly presented by Fig. 5 and Fig. 6, Informer has the best performance not only when the stock price is stable but when it is highly volatile.

For a further analysis of the performance quality of the compared networks, the more detailed evaluation criteria are enumerated in Table 3.

\begin{table}[ht]
\begin{center}
\begin{minipage}{\textwidth}
\caption{Evaluation criteria of different networks prediction performance}\label{tab3}%
\begin{tabular}{cccccc}
\toprule
Dataset & Time scale & Networks & MAE & RMSE & MAPE \\
\midrule
\multirow{8}{*}{HSI} & \multirow{4}{*}{1-minute} & LSTM & 0.048450496 & 0.07953338 & 0.2013584 \\
\multirow{8}{*}{ } & \multirow{4}{*}{ } & Transformer & 0.03281819 & 0.05688318 & 0.14695859 \\
\multirow{8}{*}{ } & \multirow{4}{*}{ } & BERT & 0.027195446 & 0.047723413 & 0.12402281 \\
\multirow{8}{*}{ } & \multirow{4}{*}{ } & Informer & 0.017365202 & 0.030502869 & 0.07789564 \\
\cline{2-6}
\multirow{8}{*}{ } & \multirow{4}{*}{5-minute} & LSTM & 0.15383348 & 0.2084857 & 0.33645868 \\
\multirow{8}{*}{ } & \multirow{4}{*}{ } & Transformer & 0.14019528 & 0.19190149 & 0.2828206 \\
\multirow{8}{*}{ } & \multirow{4}{*}{ } & BERT & 0.09063706 & 0.12669964 & 0.2349578 \\
\multirow{8}{*}{ } & \multirow{4}{*}{ } & Informer & 0.09478245 & 0.1303974 & 0.20394623 \\
\cline{1-6}
\multirow{8}{*}{IXIC} & \multirow{4}{*}{1-minute} & LSTM & 0.088538 & 0.12838154 & 0.90201306 \\
\multirow{8}{*}{ } & \multirow{4}{*}{ } & Transformer & 0.059976757 & 0.0891262 & 0.620397 \\
\multirow{8}{*}{ } & \multirow{4}{*}{ } & BERT & 0.05045773 & 0.076477446 & 0.49518508 \\
\multirow{8}{*}{ } & \multirow{4}{*}{ } & Informer & 0.033283725 & 0.04986665 & 0.32251728 \\
\cline{2-6}
\multirow{8}{*}{ } & \multirow{4}{*}{5-minute} & LSTM & 0.103707075 & 0.1373586 & 0.7210161 \\
\multirow{8}{*}{ } & \multirow{4}{*}{ } & Transformer & 0.0691121 & 0.09679416 & 0.4188139 \\
\multirow{8}{*}{ } & \multirow{4}{*}{ } & BERT & 0.05922901 & 0.08125483 & 0.3575687 \\
\multirow{8}{*}{ } & \multirow{4}{*}{ } & Informer & 0.034303054 & 0.048323717 & 0.23767516 \\
\cline{1-6}
\multirow{8}{*}{Tencent} & \multirow{4}{*}{1-minute} & LSTM & 0.055903763 & 0.08148885 & 0.25073105 \\
\multirow{8}{*}{ } & \multirow{4}{*}{ } & Transformer & 0.045478255 & 0.06317818 & 0.19847631 \\
\multirow{8}{*}{ } & \multirow{4}{*}{ } & BERT & 0.035496254 & 0.052147124 & 0.1673891 \\
\multirow{8}{*}{ } & \multirow{4}{*}{ } & Informer & 0.022001078 & 0.032871705 & 0.11511085 \\
\cline{2-6}
\multirow{8}{*}{ } & \multirow{4}{*}{5-minute} & LSTM & 0.07562126 & 0.10009563 & 0.30505136 \\
\multirow{8}{*}{ } & \multirow{4}{*}{ } & Transformer & 0.057828076 & 0.07817289 & 0.24090363 \\
\multirow{8}{*}{ } & \multirow{4}{*}{ } & BERT & 0.062069204 & 0.07904721 & 0.22131862 \\
\multirow{8}{*}{ } & \multirow{4}{*}{ } & Informer & 0.047507398 & 0.058003407 & 0.16834548 \\
\cline{1-6}
\multirow{8}{*}{AAPL} & \multirow{4}{*}{1-minute} & LSTM & 0.039807707 & 0.06286954 & 0.5211028 \\
\multirow{8}{*}{ } & \multirow{4}{*}{ } & Transformer & 0.02774789 & 0.044707693 & 0.35389292 \\
\multirow{8}{*}{ } & \multirow{4}{*}{ } & BERT & 0.022931112 & 0.037901502 & 0.300096 \\
\multirow{8}{*}{ } & \multirow{4}{*}{ } & Informer & 0.01339805 & 0.021804703 & 0.16689034 \\
\cline{2-6}
\multirow{8}{*}{ } & \multirow{4}{*}{5-minute} & LSTM & 0.11125406 & 0.15878163 & 1.2568895 \\
\multirow{8}{*}{ } & \multirow{4}{*}{ } & Transformer & 0.07619887 & 0.11191886 & 0.81891525 \\
\multirow{8}{*}{ } & \multirow{4}{*}{ } & BERT & 0.061551154 & 0.09180315 & 0.7753588 \\
\multirow{8}{*}{ } & \multirow{4}{*}{ } & Informer & 0.032591768 & 0.04985172 & 0.37670088 \\
\botrule
\end{tabular}
\footnotetext{Note: Dropout and early stopping patience mechanism is to prevent overfitting; early stopping is controlled by validation loss; learning rate is dynamic, gradually growing smaller with the training process.}
\end{minipage}
\end{center}
\end{table}

Different time scales of different stocks and indices are evaluated using the same evaluation criteria based on whole testing sets. In Table 3, it can be obtained that the prediction results of Informer outperform the prediction results of the other three networks on all three criteria, MAE, RMSE and MAPE. It’s worth mentioning that in the HSI 5-minute prediction task, the MAE and RMSE values of Informer (MAE: 0.09478245, RMSE: 0.1303974) are slightly higher than BERT (MAE: 0.09063706, RMSE: 0.12669964), but the MAPE of Informer (0.20394623) is notably lower than BERT (0.2349578), which implies that the Informer’s relative error of the predicted values is lower. According to section 4.2, Informer in this experiment is still superior to BERT.

In addition, for different data sets and different time scales, the overall results evaluation criteria are significantly different as well, which are caused by the rate of stock price volatility and the level of stock trading activity. Various stock markets, stock fundamentals and trading time determine different trading activity, leading to in the next time point, the different deviated direction and the different deviated degree of stock prices from this time. Then consider in the substantial change in volume. Obviously, the stocks with larger deviated degree, are different to be predicted, and inevitably, their results evaluation criteria turn to bad. Moreover, a shorter scale means a more inactive trading, causing a narrower range of variation, whose range in the next time point is much easier to be predicted.

Nevertheless, Informer has performed better among all the networks, especially when trading is more active and where stock prices change intensely. One of the primary reasons for this is that we take into account that in general, the most dramatic fluctuations in trading volume, the fastest price movements and the most active trading periods are during the first few minutes of the day at the opening and near the closing of the market. The global time stamp mechanism allows time to be one of the characteristics, and thereby, violent volume fluctuations instead enhance the identification ability of the model. Empirically, large trading volumes determine the dramatic fluctuations in stock prices \cite{bib40}, and price movements during the most active trading periods also determine trading sentiment and overall trends \cite{bib55} in the short term, or even throughout the day. Therefore, the daily active trading time period is of great referential value. Naturally, a good prediction of the stock price during the active period also facilitates the prediction of the stock price for the whole day. 

\subsection{Performance of global time stamp mechanism}\label{subsec4}

In the last experiment, the most important evaluation criterion is MAPE. To further research the important role played by global time stamp mechanism in the minute-level prediction task, in this section, the dataset with poor MAPE evaluation is selected for experiments. The reason why such datasets are selected is that poor MAPE means more active stocks/indices trading, which leads to greater stock price dispersion, more violent stock price fluctuations, and lower prediction accuracy. Meanwhile, it also demonstrates the positive role of the global time stamp mechanism. The MAPE values of different datasets are shown in Fig. 7, and according to the results, NASDAQ Index and Apple are selected as experimental data.

\begin{figure}[ht]%
\centering
\includegraphics[width=0.9\textwidth]{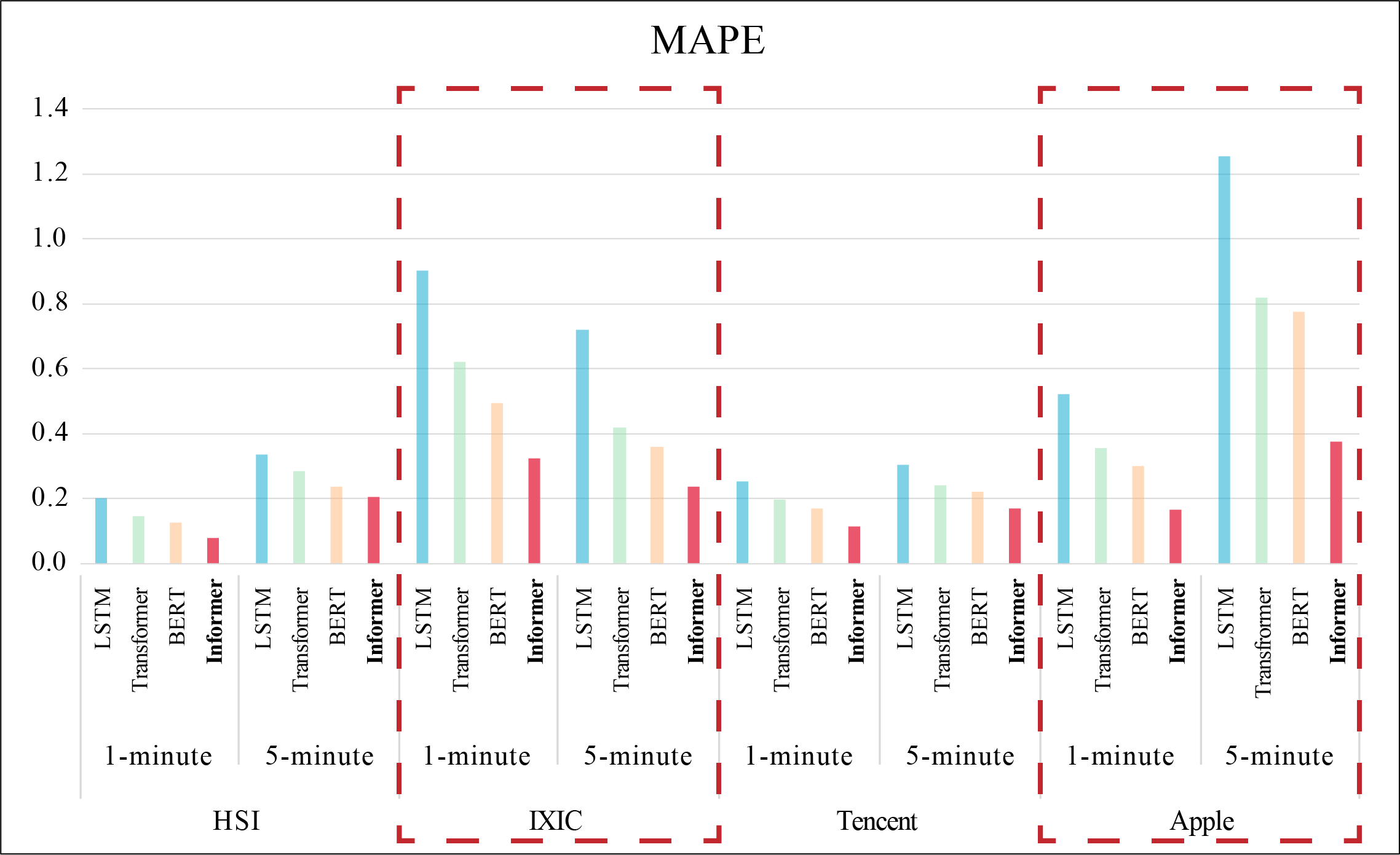}
\caption{MAPE values of compared networks on testing datasets}\label{fig7}
\end{figure}

In the new network, the global time stamp mechanism is removed and only the positional embedding in the traditional Transformer is used, and the time series is no longer considered as one of the characteristics. For the purpose of distinction, the network without the global time stamp is noted as Informer\dag in this paper. Using the same and appropriate hyper-parameters, run on selected stock/index datasets, the results are shown in the Fig. 8.

\begin{figure}[!ht]%
\centering
\includegraphics[width=0.9\textwidth]{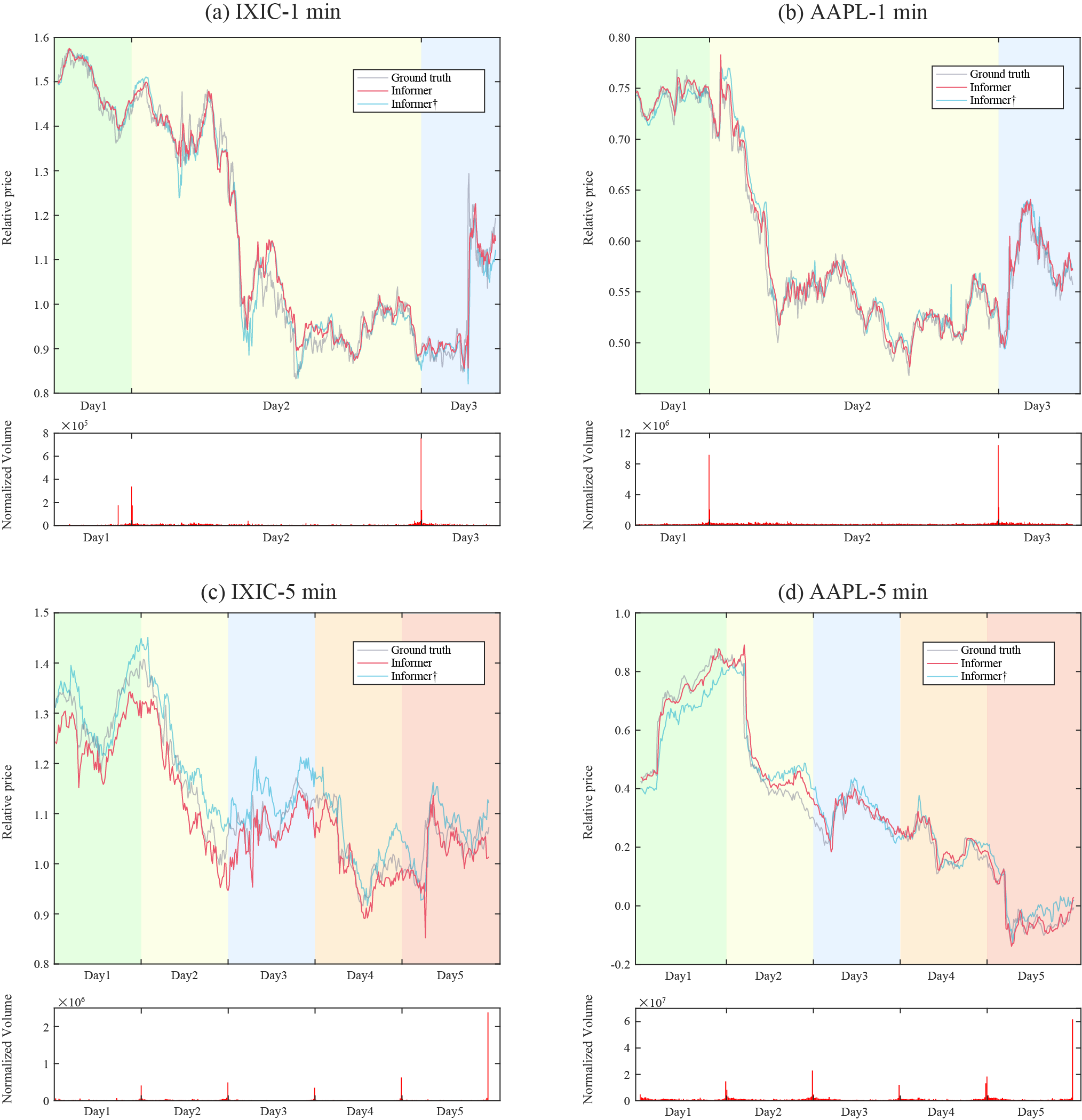}
\caption{The prediction performance of complete Informer versus Informer\dag}\label{fig8}
\end{figure}

Similarly, the figures explicitly show the predicted results with sharp local fluctuations. The results are obvious, in the period of volume spike, the stock price appears to be strongly volatile, and the prediction of Informer\dag is distinctly less effective than Informer. This well demonstrates that the global time stamp mechanism plays an important role in the network. In order to compare the experimental results more objectively, more detailed evaluation criteria values are listed in Table 4.

\begin{table}[!ht]
\begin{center}
\begin{minipage}{\textwidth}
\caption{Evaluation criteria for the complete Informer and Informer\dag}\label{tab4}%
\resizebox{\linewidth}{!}{
\begin{tabular}{cccccc}
\toprule
Dataset & Time scale  & Networks & MAE & RMSE & MAPE \\
\midrule
\multirow{4}{*}{IXIC} & \multirow{2}{*}{1-minute} & Informer & 0.033283725 & 0.04986665 & 0.32251728 \\
\multirow{4}{*}{ } & \multirow{2}{*}{ } & Informer\dag & 0.04050859 & 0.058049884 & 0.3658304 \\
\cline{2-6}
\multirow{4}{*}{ } & \multirow{2}{*}{5-minute} & Informer & 0.034303054 & 0.048323717 & 0.23767516 \\
\multirow{4}{*}{ } & \multirow{2}{*}{ } & Informer\dag & 0.045587055 & 0.06255339 & 0.32544637 \\
\cline{1-6}
\multirow{4}{*}{AAPL} & \multirow{2}{*}{1-minute} & Informer & 0.01339805 & 0.021804703 & 0.16689034 \\
\multirow{4}{*}{ } & \multirow{2}{*}{ } & Informer\dag & 0.01725388 & 0.026341373 & 0.22904986 \\
\cline{2-6}
\multirow{4}{*}{ } & \multirow{2}{*}{5-minute} & Informer & 0.032591768 & 0.04985172 & 0.37670088 \\
\multirow{4}{*}{ } & \multirow{2}{*}{ } & Informer\dag & 0.043216016 & 0.061931413 & 0.48299044 \\
\botrule
\end{tabular}
}
\footnotetext{Note: Dropout and early stopping patience mechanism is to prevent overfitting; early stopping is controlled by validation loss; learning rate is dynamic, gradually growing smaller with the training process.}
\end{minipage}
\end{center}
\end{table}

Combining Fig. 8 with Table 4, complete Informer network greatly outperforms Informer\dag, both in general and in the active parts. It can be observed that the global time stamp mechanism is very effective in intra-day stock price prediction tasks with highly volatile volumes on minute scale.

~\\

\subsection{Ability of transfer learning}\label{subsec5}

In real prediction and trading tasks, different stocks and indices have different fundamentals and intrinsic properties, which determine their trading styles. For example, there are stocks with dense pending orders and active trading, and there are also stocks with sparse pending orders, inactive trading and large stock price steps. However, it is impossible to use all stocks and indices as training sets for the neural networks to learn, which requires the neural networks to have the ability of transfer learning. That is, to predict unlearned stocks and indices based on the change characteristics of learned stocks and indices.

Therefore, in this section, we load the pre-trained weights of AAPL 1-minute dataset into Informer and use them to predict the stock prices of 1-minute and 5-minute scales of HSI, IXIC, and Tencent. This transfer learning task changes the market environments (from American market to Hong Kong market), price categories (from company stock to market index) and time scale (from 1-minute scale to 5-minute scale) between the training and testing datasets. This operation alters the character of the predicted objects, as well as the styles and features of the price movements. The pre-trained network used for transfer learning task is noted as Informer$^T$, And the ground truth, Informer predicted, and Informer$^T$ predicted performance are shown in Fig. 9.

\begin{figure}[!t]
\centering
\includegraphics[width=0.9\textwidth]{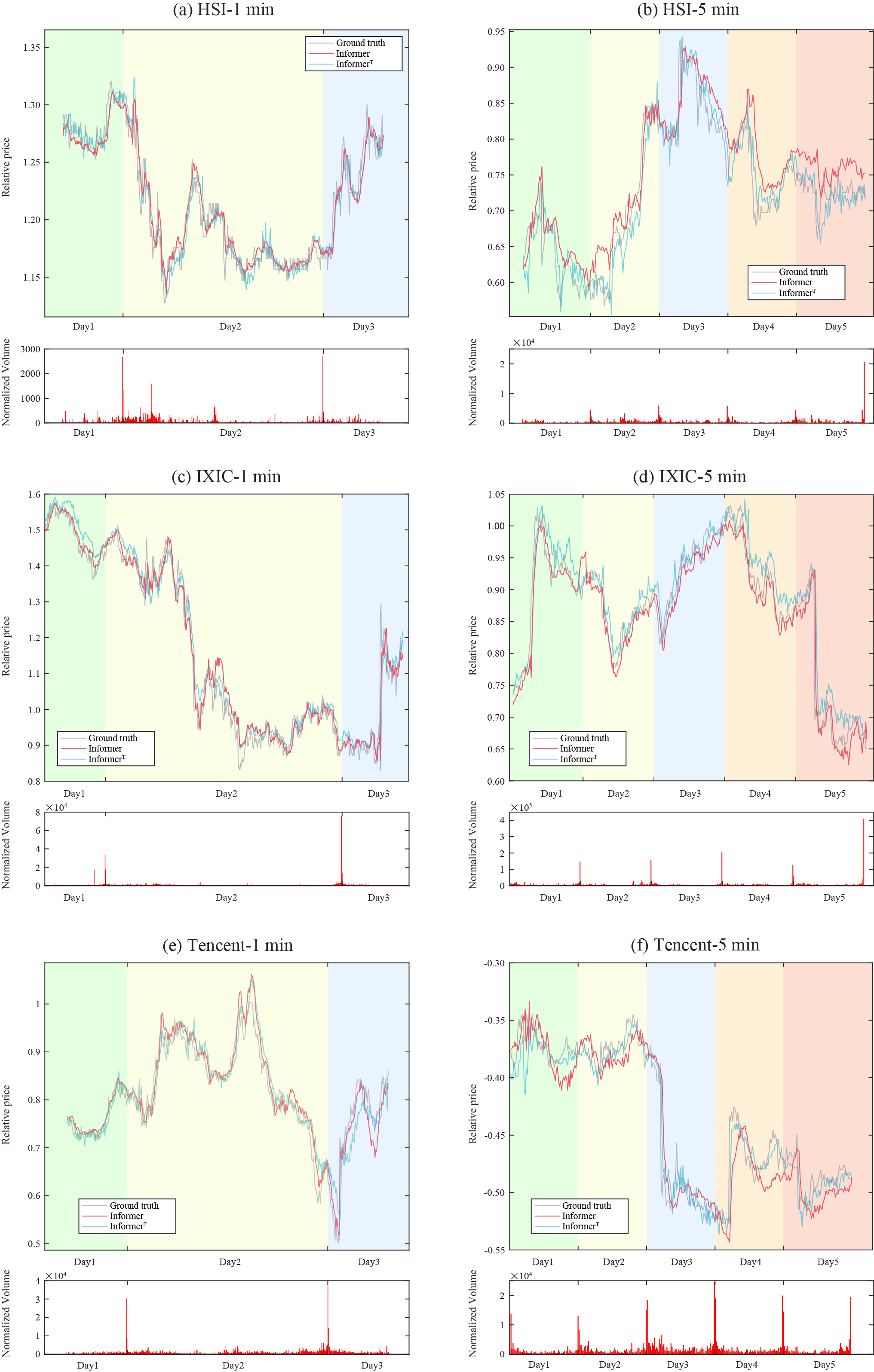}
\caption{1-minute and 5-minute scales of stock prices of HSI, IXIC, and Tencent}\label{fig9}
\end{figure}

Certain time periods are still randomly selected as the demonstration samples. Fig. 9 can display that although it does not perform as well as Informer in some periods, Informer$^T$ also has good sensitivity to the dramatically fluctuating stock prices as the trading volume changes abruptly. It still demonstrates better prediction capability than LSTM, Transformer and BERT, with less deviation and less noise from the ground truth values. Detailed evaluation criteria for transfer learning performance of Informer$^T$ is shown in Table 5. 

\begin{table}[ht]
\begin{center}
\begin{minipage}{\textwidth}
\caption{Evaluation criteria for Informer and pre-trained Informer$^T$}\label{tab5}%
\resizebox{\linewidth}{!}{
\begin{tabular}{cccccc}
\toprule
Dataset & Time scale  & Networks & MAE & RMSE & MAPE\\
\midrule
\multirow{4}{*}{HSI} & \multirow{2}{*}{1-minute} & Informer & 0.017365202 & 0.030502869 & 0.07789564 \\
\multirow{4}{*}{ } & \multirow{2}{*}{ } & Informer$^T$ & 0.020444226 & 0.036466226 & 0.08507095 \\
\cline{2-6}
\multirow{4}{*}{ } & \multirow{2}{*}{5-minute} & Informer & 0.09478245 & 0.1303974 & 0.20394623 \\
\multirow{4}{*}{ } & \multirow{2}{*}{ } & Informer$^T$ & 0.098595868 & 0.07794824 & 0.2475977 \\
\cline{1-6}
\multirow{4}{*}{IXIC} & \multirow{2}{*}{1-minute} & Informer & 0.043283725 & 0.04986665 & 0.32251728 \\
\multirow{4}{*}{ } & \multirow{2}{*}{ } & Informer$^T$ & 0.03947691 & 0.05990647 & 0.3873423 \\
\cline{2-6}
\multirow{4}{*}{ } & \multirow{2}{*}{5-minute} & Informer & 0.034303054 & 0.048323717 & 0.23767516 \\
\multirow{4}{*}{ } & \multirow{2}{*}{ } & Informer$^T$ & 0.04431727 & 0.062440593 & 0.30780962 \\
\cline{1-6}
\multirow{4}{*}{Tencent} & \multirow{2}{*}{1-minute} & Informer & 0.022001078 & 0.032871705 & 0.11511085 \\
\multirow{4}{*}{ } & \multirow{2}{*}{ } & Informer$^T$ & 0.02961939 & 0.042175338 & 0.1582657 \\
\cline{2-6}
\multirow{4}{*}{ } & \multirow{2}{*}{5-minute} & Informer & 0.047507398 & 0.058003407 & 0.16834548 \\
\multirow{4}{*}{ } & \multirow{2}{*}{ } & Informer$^T$ & 0.049185546 & 0.052125817 & 0.17809171 \\
\botrule
\end{tabular}
}
\footnotetext{Note: Dropout and early stopping patience mechanism is to prevent overfitting; early stopping is controlled by validation loss; learning rate is dynamic, gradually growing smaller with the training process.}
\end{minipage}
\end{center}
\end{table}

In the experiments of transfer learning, MAE and RMSE occasionally appear to be larger for Informer than Informer$^T$; however, in general, and according to the values of MAPE, transfer learning with different training sets still affects the prediction effect of Informer network to some extent. This is because by nature, stocks/indices in different markets have different trading rules and different participants. This fundamentally contributes to the diversity of variation and difficulty in predicting. The less correlated the two underlying are, the more difficult it will be for transfer learning. But this does not deny the excellent transfer learning ability of Informer; compared with other mature financial prediction networks, Informer has the same excellent robustness and generalizability on top of its better prediction performance.

~\\
~\\
~\\
~\\

\section{Conclusion}\label{sec4}

In technical analysis, stock movement prediction is a complex task because of the highly stochastic, highly dynamic and interconnected nature of financial markets. In this paper, the Informer network is used for the first time to predict the volatility of stock and market indices at the minute level. The major issue addressed is that the minute-scale trading volume fluctuates drastically and the rate of change is not uniform, resulting in a large relative change rate in the stock price such that the general neural network is insensitive or the results are divergent. Three experiments are designed to demonstrate the feasibility and superiority of Informer in solving this type of problem.

In our experiments on minute scale stock price prediction, we compared Informer with the commonly used LSTM, Transformer and BERT networks, set appropriate hyper-parameters, and predict the 1-minute and 5-minute frequencies of Hong Kong Hang Seng Index (HSI), NASDAQ Index (IXIC), Tencent Holdings Ltd stock and Apple Inc. (AAPL) stock separately. The results demonstrate that Informer achieved the best performance in evaluation criteria including MAE, RMSE and MAPE, compared to the other networks, and the rest compared networks prediction performance is BERT>Transformer>LSTM. These three compared networks deviate to varying degrees of great deviation when the volume changes dramatically and the stock price fluctuates sharply just after the opening and near the closing. Besides, when volume and stock price changes tend to moderate, their predictions are noisy again. In contrast, Informer exhibits the best prediction accuracy, both in localized periods with dramatic changes as well as in overall criteria measures.

To further prove the importance of the global time stamp mechanism in Informer, Experiment 2 has been conducted. The four most difficult to predict datasets from Experiment 1 are selected. The global time stamp mechanism is removed, and the network is noted as Informer\dag, compared with complete Informer with same hyper-parameters and both fully convergent. The stock price figures clearly demonstrate the apparent error of Informer\dag in the highly volatile period and the significant deterioration of the evaluation criteria score. This is because the global time stamp mechanism gives characteristics to the timing information, and the network also learns for the short to medium term stock price movements.

In practical applications, since we cannot train networks with all stock information, good transfer learning capability and robustness are equally indispensable. In Experiment 3, Informer is pre-trained only with AAPL 1-minute dataset, to predict both 1-minute and 5-minute frequencies of HSI, IXIC and Tencent. Although the results do not perform as well as the separately trained network, the stock price figures and criteria table still show that the pre-trained Informer has good sensitivity and high accuracy to the dramatic fluctuations of stock prices. The results of transfer learning are also acceptable in practical trading.

In addition, differences in markets, trading rules and stock fundamentals create very different stock trading styles, which likewise pose difficulties and challenges for the task of accurately predicting stock prices. This requires feeding the networks with a training set containing more comprehensive information and a larger scale of data volume, which also brings the problem of computing power. Therefore, how to fully obtain valuable stock information; pre-process stock information on the basis of retaining the original features; and optimize the network structure to reduce the computational complexity are still urgent problems to be solved. It is believed that with the development of artificial intelligence technology and data science, such problems can be better handled soon.

\section*{Declarations}

\textbf{Conflict of Interest} The authors declare that they have no known competing financial interests or personal relationships that could have appeared to influence the work reported in this paper.

~\\

\noindent \textbf{Data Availability} Data used to support the findings of this study are available from Tongdaxin Financial Terminal, Shenzhen Fortune Trend Technology Co., Ltd, https://www.tdx.com.cn/.

~\\

\noindent \textbf{Code Availability} Codes used to support the findings of this study are available from the corresponding author upon request.

~\\

\bibliography{article}

\end{document}